\documentclass[aps,twocolumn,superscriptaddress,showpacs,amssymb,prb,floatfix,preprintnumbers]{revtex4-1}
\usepackage{graphicx} 
\usepackage{amsmath}
\usepackage{amssymb}
\usepackage[dvipdfm]{hyperref}

\hypersetup{colorlinks=true,linkcolor=blue,citecolor=blue,urlcolor=blue}

\newcommand{\BFCoA}{Ba(Fe$_{1-x}$Co$_x$)$_2$As$_2$}
\newcommand{\BFCuA}{Ba(Fe$_{1-x}$Cu$_x$)$_2$As$_2$}
\newcommand{\BFNiA}{Ba(Fe$_{1-x}$Ni$_x$)$_2$As$_2$}
\newcommand{\BFRhA}{Ba(Fe$_{1-x}$Rh$_x$)$_2$As$_2$}
\newcommand{\BFPA}{BaFe$_2$(As$_{1-x}$P$_{x}$)$_{2}$}
\newcommand{\LFOA}{LaFeAsO$_{1-x}$F$_{x}$}
\newcommand{\slrr}{$T_1^{-1}$~}

\begin{document}

\thispagestyle{myheadings}

\title{NMR evidence for inhomogeneous glassy behavior driven by nematic fluctuations in iron arsenide superconductors}

\author{A. P. Dioguardi}
\author{M. M. Lawson}
\author{B. T. Bush}
\author{J. Crocker}
\author{K. R. Shirer}
\author{D. M. Nisson}
\author{T. Kissikov}
\affiliation{Department of Physics, University of California, Davis, California 95616, USA}
\author{S. Ran}
\author{S. L. Bud'ko}
\author{P. C. Canfield}
\affiliation{Ames Laboratory U.S. DOE and Department of Physics and Astronomy, Iowa State University, Ames, Iowa 50011, USA}
\author{S. Yuan }
\author{P. L. Kuhns}
\author{A. P. Reyes}
\affiliation{National High Magnetic Field Laboratory, Florida State University, Tallahassee, Florida 32310, USA}
\author{H.-J. Grafe}
\affiliation{IFW Dresden, Institute for Solid State Research, P.O. Box 270116, D-01171 Dresden, Germany}
\author{N. J. Curro}
\affiliation{Department of Physics, University of California, Davis, California 95616, USA}
\email{adioguardi@ucdavis.edu}

\date{\today}
\begin{abstract}

We present $^{75}$As nuclear magnetic resonance spin-lattice and spin-spin relaxation rate data in \BFCoA\ and \BFCuA\ as a function of temperature, doping and magnetic field.  The relaxation curves exhibit a broad distribution of relaxation rates, consistent with inhomogeneous glassy behavior up to 100 K. The doping and temperature response of the width of the dynamical heterogeneity is similar to that of the nematic susceptibility measured by elastoresistance measurements. We argue that quenched random fields which couple to the nematic order give rise to a nematic glass that is reflected in the spin dynamics.
\end{abstract}


\maketitle

\section{Introduction}

The iron arsenide superconductors exhibit multiple phase transitions upon doping, including antiferromagnetism, unconventional superconductivity, and electronically-driven nematic ordering that breaks $C_4$ rotation symmetry.\cite{doping122review}  In the context of  crystalline materials, nematic order refers to an orthorhombic lattice distortion that is driven by electronic rather than structural degrees of freedom.\cite{FradkinNematicReview} In the iron pnictides, the transport anisotropy far exceeds the orthorhombicity, suggesting that the origin is electronic.\cite{IronArsenideDetwinnedFisherScience2010}  The orthorhombic, or nematic, phase is characterized by the presence of perpendicular twin domains.\cite{prozorovDW122}  Importantly, there is a strong  coupling between the spin and orbital degrees of freedom ensuring that the antiferromagnetically ordered Fe spins lie along either of these two orthogonal directions.\cite{Fernandes2012}   Upon doping, the nematic and antiferromagnetic ordering temperatures are suppressed, yet strong antiferromagnetic  fluctuations persist in the paramagnetic state beyond optimal doping, even in the absence of long range order.\cite{imaiBa122overdoped}  Direct transport measurements of the electronic nematicity versus strain  have uncovered a divergent nematic susceptibility in the paramagnetic phase.\cite{FisherScienceNematic2012} The large nematic susceptibility necessarily implies the presence of nematic fluctuations in the disordered paramagnetic phase.


Nuclear magnetic resonance (NMR) has played a central role in the investigation of spin fluctuations in the iron arsenide superconductors.  The $^{75}$As nuclei ($I=3/2$, 100\% abundant) experience a strong hyperfine coupling to the neighboring Fe spins,
\cite{takigawa2008}
 thus the spin lattice relaxation rate, \slrr, is a sensitive probe of the dynamical spin susceptibility of the Fe spins.
\cite{T1formfactorsArsenides}
In the paramagnetic state of a homogeneous material, critical spin fluctuations exhibit  a characteristic time scale, $\tau_c$, that diverges as a power law at the phase transition temperature, $\tau_c \propto (T-T_N)^{-\alpha}$.  Consequently, the NMR relaxation rate $T_1^{-1} \propto \tau_c$ exhibits a sharp divergence at $T_N$. NMR studies of \slrr\ in \BFCoA\ and \BFPA\ revealed the presence of spin fluctuations over a broad range of doping and temperature, with a quantum phase transition at a critical doping level, $x_c$, that lies close to the maximal $T_c$.\cite{imaiBa122overdoped,NakaiPdopedBa122PRL,NingPRB2014}

Several recent experimental studies have reported a deviation from the expected power law divergence of \slrr\
as well as stretched exponential behavior. In \LFOA, \BFRhA, and \BFCoA, the characteristic time scale of the antiferromagnetic fluctuations grows progressively slower over a broad temperature range, the spin-lattice recovery function exhibits stretched exponential behavior, and the NMR signal intensity is suppressed (wipeout). \cite{Hammerath2013,Bossoni2013,Ba122ClusterGlassNMR,HajoPRB2014} In the case of \BFCoA, \slrr\ also  changes character in the nematic state, diverging with a critical exponent $\delta \sim \frac{1}{3}$. \cite{NingPRB2014} NMR studies at the $^{59}$Co site reveal much weaker spin fluctuations near the magnetic transition, \cite{NingJPSJ2008,ImaiLightlyDoped} and $^{63}$Cu site-selective NMR shows a similar local suppression of the spin fluctuations on the $^{63}$Cu site and neighboring $^{75}$As sites in addition to wipeout the NMR signal.\cite{Takeda:2014ia}

\begin{figure} 
	\centering
	\includegraphics[width=\linewidth]{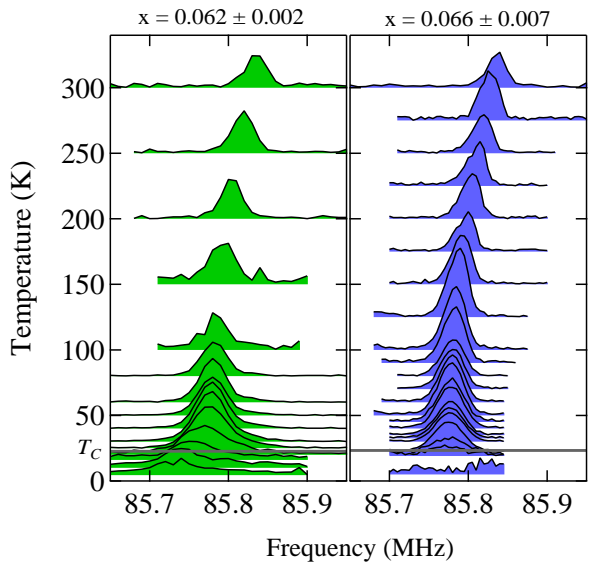}
	\caption{$^{75}$As spectra versus temperature for two different doping levels in \BFCoA\ measured by sweeping frequency at a constant field of 11.7 T and acquiring echoes for the field oriented perpendicular to the $c$ axis.  The spectra have been normalized to have equal intensities for comparison. }
	\label{fig:spectra}
\end{figure}

These features point to dynamical inhomogeneity, a characteristic of disordered spin glasses indicative of a distribution of relaxation rates, in which some fraction of the nuclei relax too quickly to be observed.
\cite{johnstonstretched,Curro2000b}
Similar behavior has been observed in the cluster spin-glass phase of the underdoped high $T_c$ cuprates,
\cite{JulienGlassyStripes2001PRB,ImaiCuprateWipeoutPRL,MitrovicGlassy214PRB2008}
and charge ordering was discovered to be intimately related to the $^{63}$Cu and $^{139}$La NQR wipeout in the cuprates.\cite{ImaiCuprateWipeoutPRL,Hunt:2001fp}
The cuprates, however,  are doped Mott insulators, and the glassy behavior was attributed to intrinsic frustration between the competing effects of Coulomb repulsion and charge segregation.
\cite{schmalianglass,WestfahlStripeGlassPRB2001}
The iron arsenides do not exhibit charge ordering and thus a different mechanism must be driving the glassy dynamics. In order to investigate the glassy behavior in more detail, we have conducted detailed field, temperature and doping dependent studies of both the spin-lattice relaxation rate, \slrr, and the spin-spin decoherence rate, $T_{2}^{-1}$.  We extract the temperature dependence of the correlation time, $\tau_c$, and find that it can be described by Vogel-Fulcher behavior. We argue that the dynamical heterogeneity arises because the dopants introduce quenched random fields coupling to the nematic order.  This disorder-induced frustration plays a significant role in suppressing antiferromagnetism and in the emergence of superconductivity.

\begin{figure} 
	\centering
	\includegraphics[width=\linewidth]{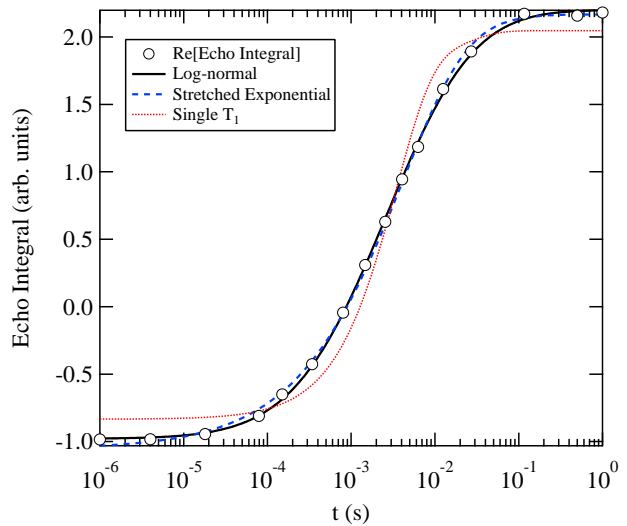}
	\caption{Magnetization versus recovery time for the $^{75}$As in \BFCoA\ with $x=0.062$ at 30K.  The solid line is the best fit using the protocol described in the text, and the dashed and dotted lines are the recovery curves assuming a stretched exponential (as described in Ref. \onlinecite{Ba122ClusterGlassNMR}) or a single relaxation time, respectively. }
	\label{fig:recovery}
\end{figure}

\section{Relaxation measurements}

Single crystals of \BFCoA\ and \BFCuA\ were grown from a FeAs self flux  and the dopant concentrations were determined via wavelength dispersive X-ray spectroscopy (WDS) as described in Ref. \onlinecite{CanfieldBa122phasediagram2008}. Multiple WDS measurements were made for each batch, and the error bars on the concentrations are given by twice the standard deviation of these measurements. $^{75}$As (100\% abundant, $I=3/2$) NMR spectra, spin lattice relaxation, and spin echo decays were measured at the central transition ($I_z = \pm 1/2$) in several different applied fields oriented perpendicular to the $c$ axis by acquiring spin echoes using standard pulse sequences. Fig. \ref{fig:spectra} shows representative spectra for two different doping levels as a function of temperature.

\begin{figure*} 
	\centering
\includegraphics[width=\linewidth]{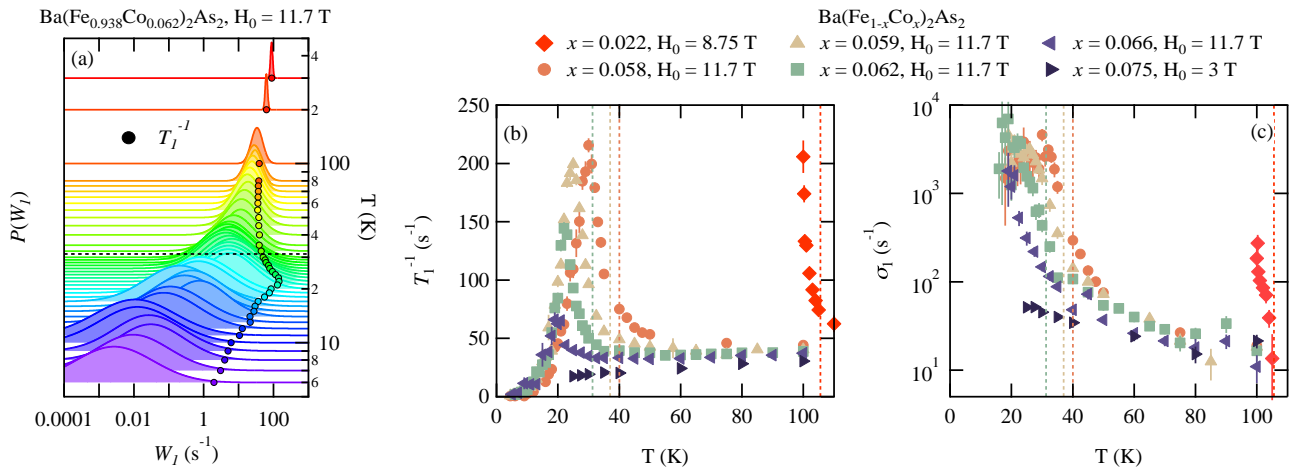}
\caption{(a) Temperature dependence of  $\mathcal{P}(W_1)$ (normalized by peak height for clarity) and the median $^{75}$As spin-lattice relaxation rate, $T_1^{-1}$, (markers) for Co-doping with $x = 0.062$ for $H_0~ ||~ ab$. Here \slrr\ $=e^{\mu}$ is the median of the distribution, $\mathcal{P}(W_1)$, as described in the text. The probability distribution broadens as temperature is decreased below $\sim100$ K. Note the bottom axis is a log scale; the skewness of the Log-Normal distribution results in the median falling on the high side of the peak (mode).  (b) \slrr\ for several Co concentrations as a function of temperature.  (c) Standard deviation $\sigma_1 = \sqrt{\langle W_1^2\rangle -\langle W_1\rangle^2}$ of the distribution $\mathcal{P}(W_1)$ for the same samples as a function of temperature in the normal state. Dashed lines in all subfigures indicate structural transition/nematic ordering temperature via bulk measurements reproduced from the literature.\cite{NSBa122incommensurate,FisherBa122PhasediagramPRB2009,CanfieldBa122phasediagram2008,NiCanfield122review,Takeda:2014ia}}
	\label{fig:distribution}
\end{figure*}

\subsection{Relaxation rate distribution}

\begin{figure} 
	\centering
	\includegraphics[width=\linewidth]{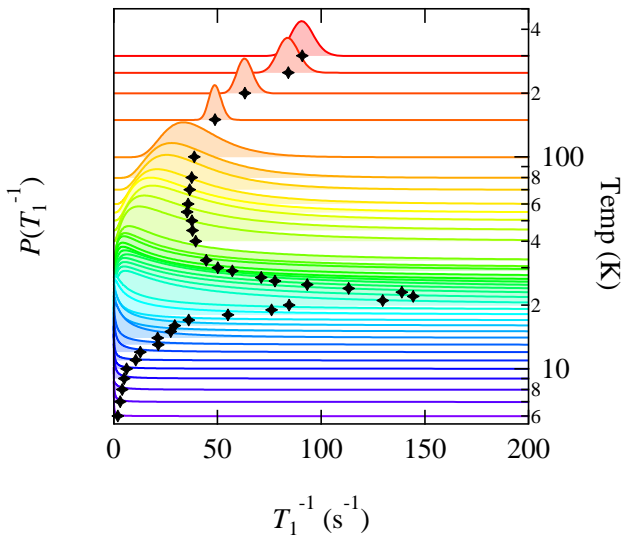}
	\caption{Temperature dependence of  $\mathcal{P}(W_1)$ (normalized by peak height for clarity) and the median $^{75}$As spin-lattice relaxation rate, $T_1^{-1}$, (markers) for Co-doping with $x = 0.062$ for $H_0~ ||~ ab$. The data is identical to that in Fig. \ref{fig:distribution}(a), but is plotted on a linear scale.  }
	\label{fig:lineardistribution}
\end{figure}

In order to quantify the distribution of relaxation rates, we fit the $^{75}$As magnetization recovery to a distribution: $M(t) = \int \mathcal{P}(W_1) f(W_1 t) d W_1$, where $\mathcal{P}(W_1)$ describes the relaxation rate distribution, and the relaxation function $f(x)$ is described below.  For a homogeneous system $\mathcal{P}(W_1)$ is a delta function centered at $T_1^{-1}$ and thus $M(t) \sim f(t/T_1)$.  If the distribution has a finite width, then the recovery function is more complex, typically exhibiting stretched behavior.  For example, if the relaxation function $f(x) = e^{-x}$, then $M(t) \sim e^{-(t/T_1)^\beta}$, where $\beta\leq 1$ is the stretching exponent.\cite{johnstonstretched}
Previous studies have reported stretched recovery, however the distribution function for general $\beta$ can only be expressed as an infinite series. Here we assume a log-normal distribution $\mathcal{P}(W_1)$ with median $T_{1}^{-1} = e^{\mu}$ and standard deviation $\sigma_1$, and fit the magnetization recovery directly. This form was chosen because it mimics the distribution for a stretched exponential recovery. This approach enables us to extract the width of the dynamical distribution of the nuclei that contribute to the NMR signal, a quantity that sheds important light on the glassy behavior.

\begin{figure*}[t!] 
	\centering
	\includegraphics[width=\textwidth]{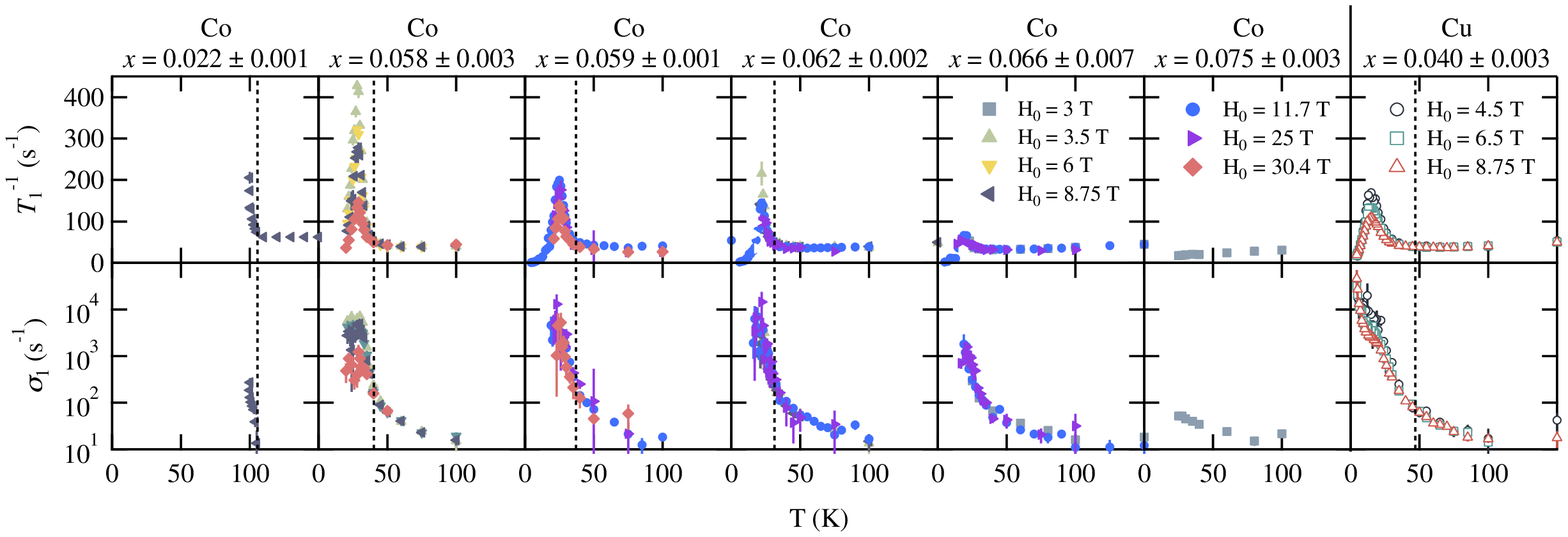}
	\caption{Field and temperature dependence of the median (\slrr) and standard deviation ($\sigma_1$) of the distribution of $^{75}$As relaxation rates for Co- and Cu-doped BaFe$_2$As$_2$. The peak in \slrr\ is strongly field dependent, typical for glassy dynamics.  $\sigma_1$ grows substantially ($\sim 10^4$ s$^{-1}$) below 100 K, reflecting the inhomogeneous relaxation of the nuclei.  Dashed lines indicate structural transition temperatures via bulk measurements reproduced from the literature.~\cite{NSBa122incommensurate,FisherBa122PhasediagramPRB2009,CanfieldBa122phasediagram2008,NiCanfield122review,Takeda:2014ia}}
	\label{fig:fielddependence}
\end{figure*}

A representative recovery data set with the best fit is shown in Fig. \ref{fig:recovery}. The distribution function is given by:
\begin{equation}
	\label{eqn:LogNormal_dist_T1}
	\mathcal{P}(W_1) = \frac{1}{W_1 \sigma \sqrt{2\pi}} e^{-\frac{1}{2}\left(\frac{\ln{W_1} - \mu}{\sigma}\right)^2},
\end{equation}
where $\mu$ and $\sigma$ are variable parameters. We define $T_1^{-1}$ as the median of the distribution, $e^{\mu}$, and the standard deviation is given by $\sigma_1 = \sqrt{e^{2\mu+\sigma^2}\left(e^{\sigma^2}-1\right)}$. The recovery function for a spin 3/2 nucleus at the central transition is:
\begin{equation}
 f(x) =M_0\left[1-2\phi\left(\frac{9}{10}e^{-6x} + \frac{1}{10}e^{-x}\right)\right],
 \label{eqn:fx}
\end{equation}
where $x=W_1t$,  $t$ is the time between the initial inverting (or saturating) pulse and the spin echo which samples the nuclear magnetization, $M_0$ is the equilibrium nuclear magnetization and $\phi$ is the inversion fraction.   $M(t)$ was numerically integrated during fitting using an adaptive Gaussian quadrature method and recalculated iteratively using a least squares method. The limits for the numerical integration were chosen to be $10^{-6}$s and $10^{6}$s, though choosing a smaller range when $\sigma$ is small results in faster convergence. This choice of limits was made based on the timescale of the NMR experiment. Spins that relax faster or slower than this time window will not participate in the spin echo, and therefore provide natural limits of integration.  The solid line in Fig. \ref{fig:recovery} shows the best fit determined in this fashion, as well as the best fits assuming either a stretched exponential, or a single value of $T_1^{-1}$ using Eq. \ref{eqn:fx}. It is clear that a single uniform relaxation rate does not accurately describe the data, but both the stretched exponential and the distribution fit  well.

Figs. \ref{fig:distribution}(a) and \ref{fig:lineardistribution} show the temperature dependence of the distribution $\mathcal{P}(W_1)$,  $T_1^{-1}$, and $\sigma_1$ as a function of temperature for \BFCoA. The data reveal a progressive broadening of the distribution below 100K, as well as an increase in both $T_{1}^{-1}$ and $\sigma_1$ reaching a peak at a temperature that coincides with the onset of long-range antiferromagnetic order at $T_N$.  The peak temperature is strongly doping dependent, reflecting the suppression of $T_N$ with doping concentration.  The  width $\sigma_1$ increases by two orders of magnitude, and is also doping dependent. This quantity is a direct measure of the degree of dynamical inhomogeneity of the system. Note that at low temperatures it is likely that the true width is even larger, but we are unable to capture the full distribution due to signal wipeout.  A previous NMR study revealed that \BFCoA\ forms a cluster spin-glass state at low temperature below $T_N$, characterized by a distribution of frozen antiferromagnetic domains coexisting with superconductivity.
\cite{Ba122ClusterGlassNMR}
Subsequent neutron scattering work concluded that this cluster spin-glass (or as termed by Lu \textit{et al.} ``moment amplitude spin glass'') state emerges also in \BFNiA.\cite{NeutronsClusterGlassBaFe2As2}
The NMR data, however indicate that this inhomogeneity begins to form  at $\sim 100$ K, well above $T_N$, where the spins are fluctuating dynamically. This large onset temperature suggests that the inhomogeneous fluctuations are unrelated to the presence of superconductivity which emerges only below $T_N$.  Furthermore, if the glassy behavior arises strictly from disorder and frustration among the spin exchange interactions, it is surprising that the inhomogeneity would emerge at temperatures well above $T_N$, where the spin presumably remain uncoupled.

\subsection{Field and doping dependence}

\begin{figure} 
	\centering
	\includegraphics[width=\linewidth]{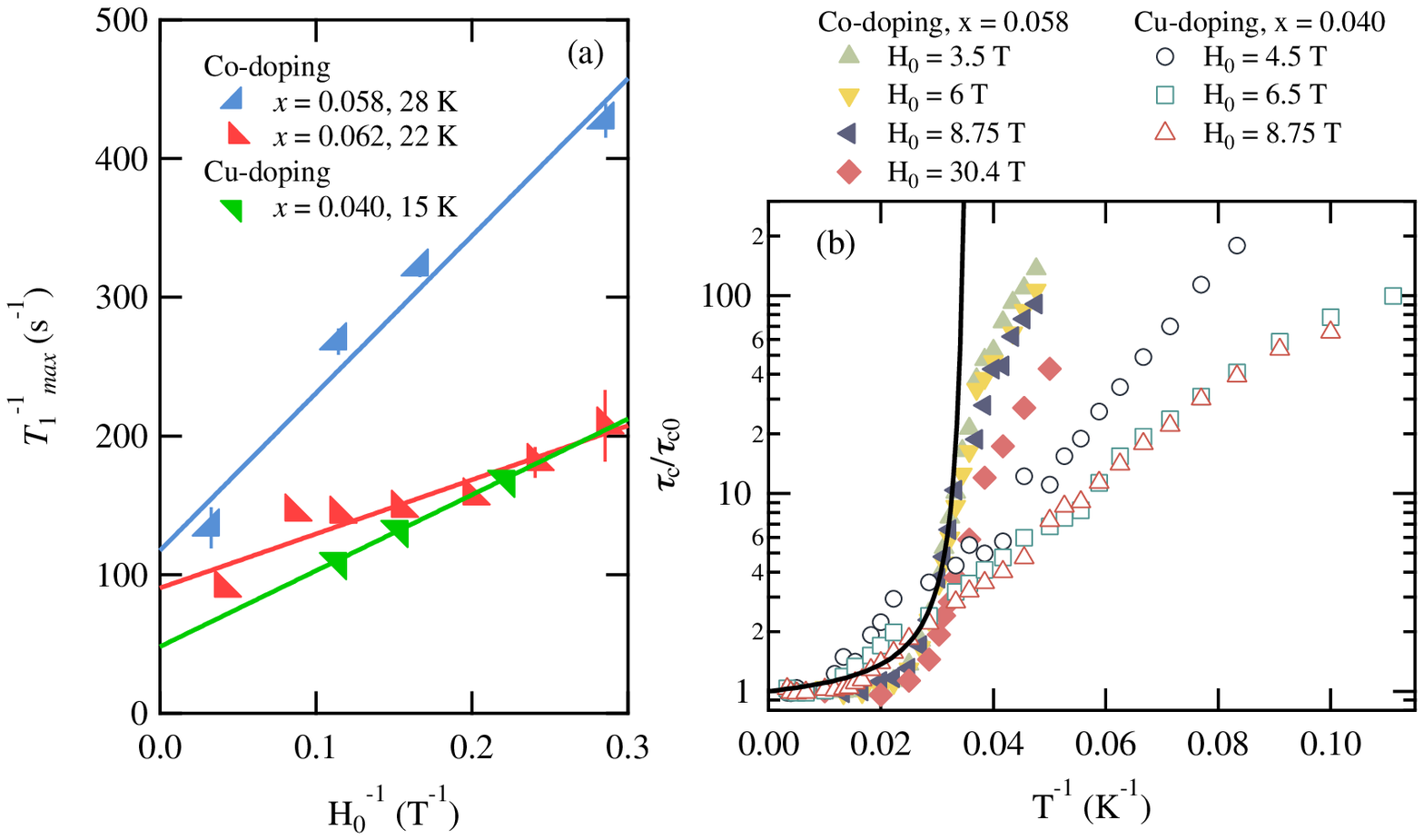}
	\caption{(a) $T_{1,max}^{-1}$ versus $H_0^{-1}$ for $x=0.058$ and $x=0.062$ in \BFCoA\ and $x=0.04$ in \BFCuA.  The slope of these data sets reveal the RMS hyperfine field values at the As site, and the fit coefficients are detailed in Table~\ref{tab:T1invmax_vs_H0inv_fitpars}. (b) Arrhenius plot of $\log(\tau_c/\tau_{c0})$ versus inverse temperature for several different fields for Co doping with $x=0.058$ and Cu doping with $x=0.04$.  The solid black line shows a Vogel-Fulcher-Tamman function, as described in the text.}
	\label{fig:taucVStemp}
\end{figure}

In order to explore the glassy behavior in more detail, we have carried out detailed studies of the field and temperature dependence of  $\mathcal{P}(W_1)$ as a function of doping in both superconducting and non-superconducting samples.  Changing the magnetic field alters the Larmor frequency, enabling one to probe the frequency dependence of the slow dynamics.  We  measured the relaxation in both \BFCoA\ (up to 30.4 T at the National High Magnetic Field Laboratory) and \BFCuA\ (up to 8.75 T). Fig. \ref{fig:fielddependence} shows $T_1^{-1}$ for several different doping concentrations and fields as a function of temperature in  \BFCoA\ and \BFCuA. Both Co and Cu dopants suppress the long range nematic and antiferromagnetic ordering, but Cu also suppresses superconductivity to a maximum $T_c \approx 2$ K, whereas $T_c$ reaches a maximum of 23 K in Co-doped samples.
\cite{doping122review,NiCanfield122review}
This enables us to discern whether the glassy behavior is connected to the competing superconducting and antiferromagnetic ground states.
\cite{NussinovPRB2009}
Both systems exhibit qualitatively similar glassy behavior, suggesting that its origin is unrelated to the superconductivity. The maximum \slrr\ is suppressed with field, reflecting the fact that the relaxation measurement is sampling the fluctuation spectrum at a different Larmor frequency. For a hyperfine field $h(t)$, the autocorrelation function is given by $\langle h(t)h(0)\rangle = h_0^2 e^{-t/\tau_c}$, where $h_0$ is the root mean square value of the field and $\tau_c$ is the autocorrelation time.\cite{CPSbook} In this case, the nuclear spin-lattice relaxation rate is:
\begin{equation}
W_1^{-1} = \frac{\gamma^2 h_0^2 \tau_c}{1 + \omega_L^2\tau_c^2},
\label{eqn:tone}
 \end{equation}
where $\gamma= 7.2919$ MHz/T is the $^{75}$As gyromagnetic ratio and $\omega_L = \gamma H_0$ is the NMR Larmor frequency.  Note that $\mathcal{P}(W_1)$ reflects a distribution of both $\tau_c$ and $h_0$.  For concreteness, however, we consider only single values of these quantities giving rise to the median of the distribution, $T_1^{-1}$, which is an oversimplification for the real system.  Eq. \ref{eqn:tone} shows that $T_1^{-1}$ reaches a maximum when $\omega_L \tau_c = 1$ and is equal to $T_{1,max}^{-1} = \gamma h_0^2/2 H_0$.  Fig. \ref{fig:taucVStemp}(a) shows $T_{1,max}^{-1}$ varies linearly with  $H_0^{-1}$ for various dopings, as expected.  The slope of this line gives $h_0$ (fit values given in Table \ref{tab:T1invmax_vs_H0inv_fitpars}), which decreases with dopant concentration, and agrees with previous measurements in LaFeAsO$_{1-x}$F$_{x}$.
\cite{Hammerath2013}

\begin{table}
	\begin{ruledtabular}
		\begin{tabular}{llll}
		Dopant 	& Doping $x$ 	& $h_0$ (mT) 			& offset (s$^{-1}$) 		\\
		\hline \\[-0.3cm]
		Co	 	& 0.058 		& 7.04 $\pm$ 0.43	& 117.32 $\pm$ 24.50 \\
		Co 	 	& 0.062 		& 4.13 $\pm$ 0.34	& 90.28 $\pm$ 11.60 	\\
		Cu	 	& 0.040 		& 4.89 $\pm$ 0.07	& 48.02 $\pm$ 2.77
		\end{tabular}
	\end{ruledtabular}
	\caption{\label{tab:T1invmax_vs_H0inv_fitpars}Fit parameters extracted for linear fits to $T_{1}^{-1,max}(H_0^{-1})$.}
\end{table}

Using the measured $h_0$, we proceed to extract $\tau_c$. Solving Eqn. \ref{eqn:tone} for $\tau_c$ yields:
\begin{equation}
\tau_c = \omega_L^{-1}\left[\frac{T_{1,max}^{-1}}{T_1^{-1}}\pm\sqrt{\left(\frac{T_{1,max}^{-1}}{T_1^{-1}}\right)^2-1}\right],
\end{equation}
where the positive sign for the radical arises at low temperature below $T_{1,max}^{-1}$ where $\tau_c\gg \omega_L^{-1}$, and the negative sign arises at high temperatures when $\tau_c\ll \omega_L^{-1}$.  Fig. \ref{fig:taucVStemp}(b) presents an Arrhenius plot of  $\tau_c/\tau_{c0}$ versus $T^{-1}$, where $\tau_{c0}$ is the high temperature limit of the correlation time.  The data clearly deviate from linearity, indicating that there is not a single activation energy that describes the system.  The solid black line represents a Vogel-Fulcher-Tamman law ($\tau_c/\tau_{c0} = \exp(DT_K/(T-T_K))$, with $D = 0.5(4)$ and $T_K = 25(3)$K).  This behavior is often found in glassy systems, and indicates  a `fragile' glass, in which the effective activation energy increases with decreasing temperature reflecting the collective nature of the fluctuations.
\cite{BerthierRMPglasses}
$T_K$ represents the temperature below which the system becomes trapped in a local minimum in free energy at a glass transition temperature. In this case, $T_K$ appears to correspond roughly with the N\'{e}el temperature. However, based on constant field Co-doping variation fits, this trend appears to break down once $T_N(\mathrm{NMR}) < T_c$, where $T_N({NMR})$ is the temperature at which \slrr reaches a maximum. Below this temperature, the spins are ordered in frozen clusters with a broad distribution of sizes.
\cite{Ba122ClusterGlassNMR,NeutronsClusterGlassBaFe2As2}
For the Cu-doped system, the $\tau_c$ exhibits more Arrhenius-type behavior. At $x=0.04$, the peak temperature of \slrr\ is $\approx 20$ K, which agrees with the phase diagram determined via bulk transport and magnetization.
\cite{NiCanfield122review}
It is unclear why the Cu-doped samples differ, but the data suggest that the fluctuations are less correlated in this system, which may, in turn, be related to the strong suppression of the superconductivity in this compound. Recent $^{63}$Cu NMR data suggest a strong local effect of the dopants, supporting such an interpretation.\cite{Takeda:2014ia}  

The data in Figs. \ref{fig:fielddependence} and \ref{fig:taucVStemp}(b) indicate that $\mathcal{P}(W_1)$ is slightly modified by the field.  In particular, the median fluctuation rate $\tau_c$  and the width $\sigma_1$ are suppressed by fields up to 30.4 T in the Co-doped sample and 8.75 T in the Cu-doped sample.
These results suggest that in high fields the distribution of domain sizes is narrowing and shifting toward smaller domains.  Note that because of the wipeout effects, these characterizations of the temperature dependence of the glassy behavior may not fully capture the behavior of the entire distribution. Since we are unable to detect large domains (with correspondingly large correlation times $\tau_c$) due to wipeout, it is possible that the field alters the domain distribution in a manner that shifts the weight of the observed distribution towards smaller sizes. Superconductivity in the Co-doped samples is also strongly suppressed in these fields, which may alter somewhat the domain distribution.
\cite{CanfieldBa122phasediagram2008}

\subsection{Spin Echo Decay}

\begin{figure*}[t!] 
	\centering
\includegraphics[width=\textwidth]{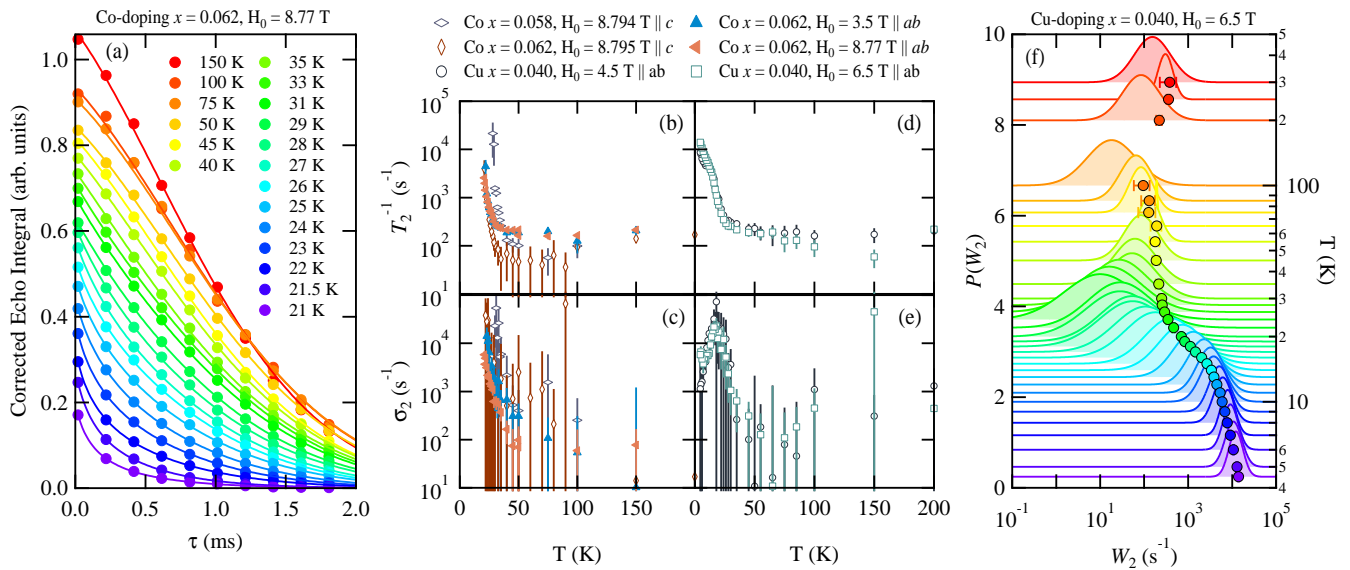}
	\caption{Temperature and field dependence of $^{75}$As echo decay for Co- and Cu-doped BaFe$_2$As$_2$. (a) Echo decay curves for $x=0.062$, scaled by the nuclear Curie susceptibility. These data are fit (solid lines, see text for details) to extract the distribution of spin-spin relaxation rates, $\mathcal{P}(W_2)$.
The data were globally fit holding the Gaussian component constant as a function of temperature, and by employing the same log-normal distribution form to fit the exponential component.
(b) The median ($T_2^{-1}$) and (c) standard deviation ($\sigma_2$) of the distribution versus temperature for several Co-doped samples.  (d) $T_2^{-1}$ (e) $\sigma_2$ versus temperature for the Cu-doped sample.  (f) The distribution $\mathcal{P}(W_2)$ and median $T_2^{-1}$ (markers) versus temperature for $x=0.04$ at 6.5 T for \BFCuA.}
	\label{fig:T2_summary}
\end{figure*}

Further evidence for glassy behavior is found in the temperature dependence of the $^{75}$As spin-echo decay curves. In addition to the increase in  $\tau_c/\tau_{c0}$ and $\sigma_1$, the NMR signal intensity gradually becomes suppressed and the character of the echo decay changes below 100 K. Fig. \ref{fig:T2_summary}(a) shows  the echo intensity following a standard echo pulse sequence ($\frac{\pi}{2}-\tau-\pi-\tau$)  for \BFCoA\ with $x=0.062$. The intensity decreases with pulse spacing $\tau$ due to various decoherence effects, including fluctuations of the hyperfine field, $h(t)$, over the course of the spin echo experiment. The data have been normalized by temperature to account for the Curie susceptibility of the nuclei, and clearly reveal the suppression of intensity (wipeout) with decreasing temperature.\cite{Ba122ClusterGlassNMR} As seen in Fig.
\ref{fig:T2_summary}(a), the character of the echo decay function crosses over from a Gaussian-dominated decay at high temperatures to exponential decay below $\sim$ 100 K.  This crossover is due to the growth of fast spin fluctuations, contributing a factor $e^{-2W_2\tau}$ to the echo decay, with $W_{2} = \gamma^2 h_z^2\tau_c$. Here $h_z^2$ is the root mean square of the hyperfine field parallel to $H_0$, in contrast to $h_0$ in Eq. \ref{eqn:tone} which lies perpendicular to $H_0$.
\cite{CPSbook}
Since there is a distribution of correlation times $\tau_c$ as evident from the \slrr\ data, we fit the echo decay data with the same protocol involving a distribution of decoherence rates, $W_2$.  The data were fit to the function: $M(2\tau) = \int_0^\infty \mathcal{P}(W_2)g(2\tau)dW_2$, where $g(\tau) =  M_0 e^{-(2\tau)^2/2 T_{2G}^2} e^{- 2 W_2 \tau}$. Here $W_2$ is the exponential component of the spin-spin relaxation rate due to spin-fluctuations, $\tau$ is the time separating the $\pi/2$ and $\pi$ pulses of the spin echo sequence, and $T_{2G}$ is the temperature independent Gaussian component of the spin-spin relaxation. At high temperatures the echo decay has a Gaussian form, which reflects the complex direct and indirect couplings between the like As nuclei.  We do not expect this component to change with temperature, whereas the growth of spin fluctuations at low temperature will affect $W_2$.
\cite{CPSbook,Curro1998}
Each temperature dependent data set was fit globally with a temperature-independent $T_{2G}$ to achieve the best fit to all temperatures. This global analysis was confirmed by individually fitting the data set at each temperature, results of which show no trend in $T_{2G}$ as a function of temperature.

The data in Fig. \ref{fig:T2_summary}(f) shows  $\mathcal{P}(W_2)$,  panels (b) and (d) show the median $T_2^{-1}$ and panels (c) and (e) show the standard deviation $\sigma_2$ for several doping levels and dopants as a function of temperature.  The temperature dependence of $T_2^{-1}$ agrees qualitatively with the correlation times extracted from the \slrr\ data seen in Fig. \ref{fig:taucVStemp}(b). $T_2^{-1} = \gamma^2 h_\parallel^2 \tau_c$, therefore we expect a monotonic increase of $T_2^{-1}$ with decreasing temperature. Surprisingly, the width $\sigma_2$ of this distribution differs from $\sigma_1$ extracted from the spin-lattice relaxation data, and exhibits a downturn below $T_N$.  Note, however, that  $\mathcal{P}(W_2)$ is cut off at large $W_2$ by the finite detection window of the NMR experiment, which is the primary cause of signal wipeout.
\cite{Curro2000b}
As a result, the measured width $\sigma_2$ is reduced as the majority of the distribution shifts outside of the detection window at low temperature.

\section{Discussion}

\subsection{Missing signal}

It is clear from Fig. \ref{fig:T2_summary} that signal wipeout of up to 80\% is present, consistent with previous measurements of the spectral intensity in these samples, which raises the question of where the missing signal has gone.\cite{Ba122ClusterGlassNMR}  The system is either dynamically or spectrally inhomogeneous.  In our experiments we find no significant broadening of the spectra in the paramagnetic state, as shown in Fig. \ref{fig:spectra}.  It is possible that the distribution is such that a large fraction of the nuclei resonate outside of this window, but the internal field in the ordered state is small in this range of dopings and the spectral shift for this field orientation is minimal.\cite{Dioguardi2010}  The spin lattice relaxation was measured at the peak of this resonance, and it is possible that not all of the nuclei were inverted by the radiofrequency pulses. 
It is more likely, however, that the missing signal arises from dynamical heterogeneity, given the broad distribution of relaxation rates that we observe. The missing signal in this case arises from nuclei that are located in an environment with a sufficiently long $\tau_c$ such that they recover to equilibrium before they can contribute to the spin-echo signal. It is important to note that the distributions shown in Fig. \ref{fig:distribution}, \ref{fig:lineardistribution} and \ref{fig:T2_summary} are representative only of the nuclei that are actually contributing to the signal. In fact the true distributions are likely to be much broader than what we are able to measure, as a significant portion of the nuclei experience even faster relaxation rates.

\subsection{Glassy nematic fluctuations}

\begin{figure} 
	\centering
\includegraphics[width=\linewidth]{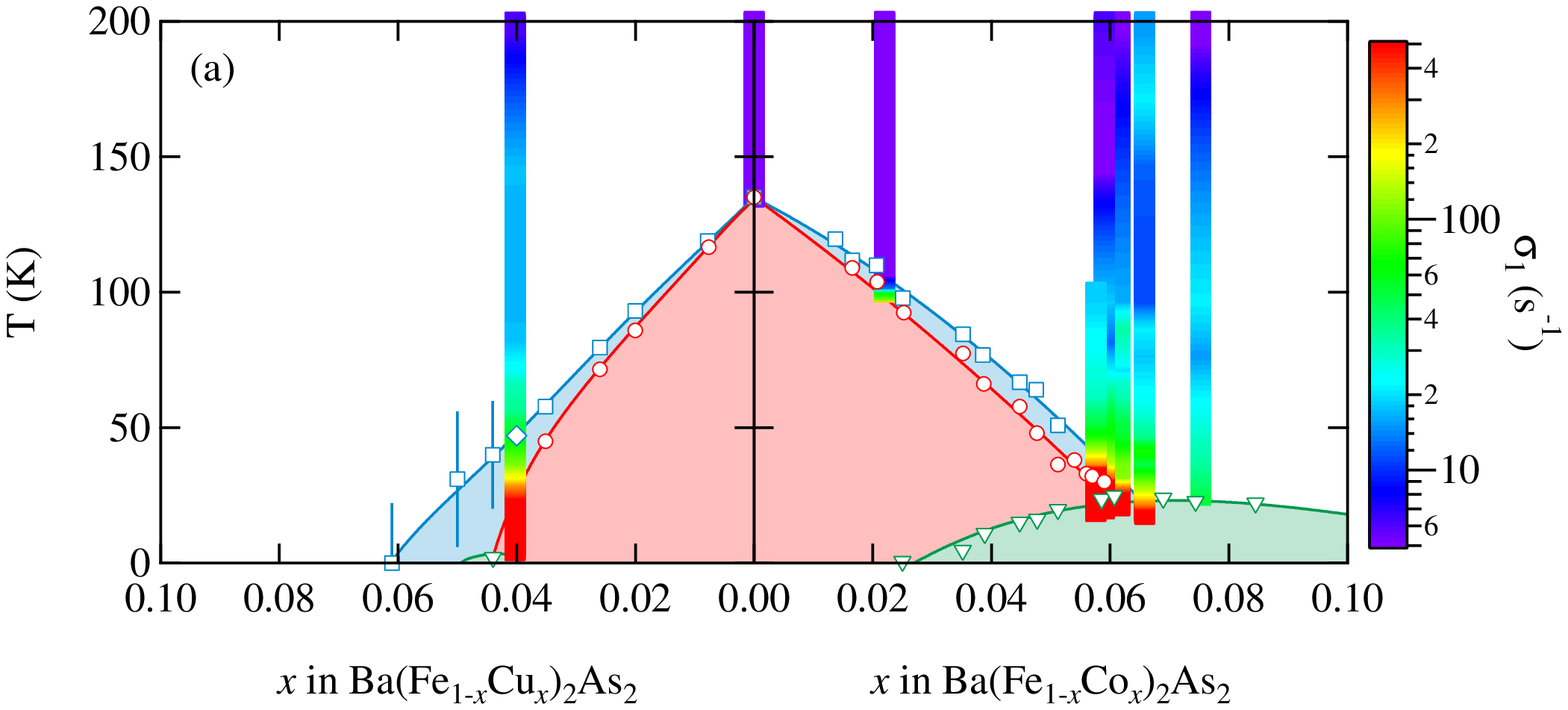}
\includegraphics[width=0.8\linewidth]{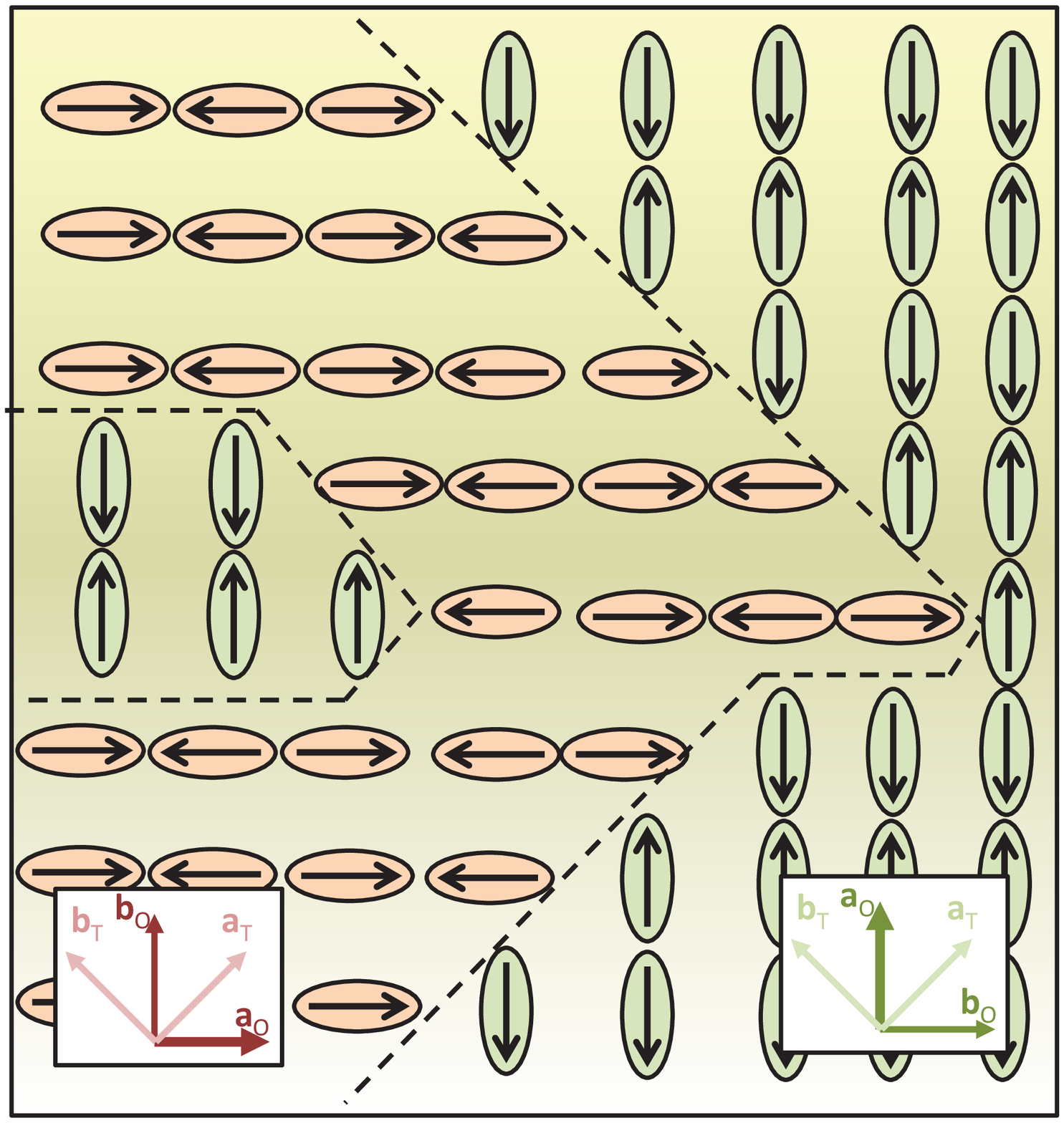}
   \caption{(a) Temperature vs. doping phase diagram of Cu- and Co-doped BaFe$_2$As$_2$. Markers have been reproduced from bulk measurements in the literature, and solid lines are a guide to the eye.~\cite{NSBa122incommensurate,FisherBa122PhasediagramPRB2009, CanfieldBa122phasediagram2008,NiCanfield122review,Takeda:2014ia} The color scale overlay shows the standard deviation $\sigma_1 = \sqrt{\langle W_1^2\rangle -\langle W_1\rangle^2}$ for the distribution $\mathcal{P}(W_1)$, characterizing the degree of inhomogeneity of the NMR  spin-lattice relaxation rate. (b) Schematic of local nematic domains, indicating directions of Fe spin (arrows).  The tetragonal and orthorhombic unit cell axes are shown.  The local nematicity is oriented along the ellipses.}
	\label{fig:phasediagram}
\end{figure}

The inhomogeneous fluctuation distribution cannot be understood simply in terms of critical slowing down of the spin degrees of freedom.  The spin fluctuations are not averaged out spatially, implying the existence of multiple local domains of characteristic size $\xi\sim\tau_c$. Figure \ref{fig:phasediagram}(a) summarizes the doping dependence of the width, $\sigma_1(x,T)$, of the inhomogeneous distribution, where $\sigma_1$ is related to the distribution of domain sizes.  For the parent compound BaFe$_2$As$_2$ we find that the recovery fits best to a single component of relaxation for all temperatures, so the system is dynamically homogeneous. $\sigma_1$ remains small for the lightly Co-doped regime; however near optimal doping it becomes a strong function of temperature, reflecting a large dynamical inhomogeneity both in the Co and Cu-doped crystals.

A likely origin for this inhomogeneity  is nematic fluctuations associated with the proximate tetragonal-to-orthorhombic structural phase transition. The doping and temperature trends exhibited by $\sigma_1(x,T)$ shown in Fig. \ref{fig:phasediagram}(a) closely resemble the behavior of the static nematic susceptibility, $\chi_n$.\cite{FisherScienceNematic2012}
\cite{IronArsenideDetwinnedFisherScience2010}
Chu et al. have found that $\chi_n(T)$ exhibits Curie-Weiss behavior, with Weiss temperature $\theta$ that vanishes at the critical doping of $x_c = 0.07$ for the Co-doped system.   The fluctuation-dissipation theorem implies that because of the large susceptibility, there are also significant thermal fluctuations of the nematic order.  In other words, even though there is no long-range nematic order, local orthorhombic distortions continue to fluctuate well above the ordering temperature.  Because the spins are strongly coupled to the nematicity, these nematic fluctuations will drive spin fluctuations, which in turn couple to the nuclei via the hyperfine interaction to influence nuclear spin-lattice relaxation.  In fact, $T_1^{-1}$ scales with shear modulus in this phase, reflecting the fact that both quantities are probing the dynamics of the nematic fluctuations.\cite{FernandesPRLnematicT1}

The glassy inhomogeneous nature of the fluctuations, therefore, probably reflects a property of the nematic fluctuations. Because the nematic order has Ising symmetry and  breaks spatial symmetry, it is highly sensitive to quenched random impurities and is prone to exhibit glassy behavior.\cite{ImryMaRandomDisorderPRL}
The theory of electronic nematic order and the role of disorder is well established in the context of the cuprates, \cite{KivelsonLiquidXtal1998,HysteresisNematicPRL,CarlsonDahmenNature2010, IsingSpinOrderSachdev2008,RFIMsimulations2010,DisorderNematicityCuprates2014}, and more recently in the context of the iron pnictides.\cite{NematicOrderLaFeAsO2008,Kuo2015}
The dopant atoms may provide a random field potential for nematic order
that suppresses the phase transition temperature and gives rise to a distribution of frustrated nematic domains with different fluctuation rates, as illustrated in Fig. \ref{fig:phasediagram}(b). With increasing dopant concentration, the nematic ordering transition is gradually suppressed. In the disordered phase, there are fluctuating patches in which $C_4$ symmetry is temporarily and locally broken, but there is no long range or static order.  These fluctuating patches, however, exhibit a broad range of sizes and fluctuation times. The inhomogeneity we observe reflects the distribution of these patches. The NMR data indicate that the nematic fluctuations and distribution of domains persist up to $\sim 100$ K, as shown in Fig. \ref{fig:phasediagram}(a).  The local autocorrelation time of the domains, $\tau_c$, is proportional to the domain size, thus the width of the distribution of domain sizes grows up to two orders of magnitude by the onset of long range nematic order.  This scenario provides a natural explanation for the large $\chi_n$  as well as the broad distribution of relaxation times observed in our NMR experiments. Further, it explains the similarity of the phase diagram of both electron and hole-doped systems, as well as the isovalent \BFPA\ system.
\cite{NakaiPdopedBa122PRL}

The temperature-pressure phase diagram of the stoichiometric parent compound also exhibits a suppression of antiferromagnetism and emergence of superconductivity without the presence of dopants;
\cite{A122phasediagramPressure,Colombier2010}
however, natural lattice defects may provide a source of quenched disorder that could be amplified by non-hydrostatic pressure. On the other hand, non-isovalent dopants clearly play a role in tuning the density of states, as revealed by a recent study of simultaneous hole- and electron-doping in Ba$_{1−x}$K$_x$Fe$_{1.86}$Co$_{0.14}$As$_2$ demonstrating that that the magnetic state can be partially recovered by compensating the carrier concentration.
\cite{Ba122codoping}
Thus both disorder and tuning the density of states appear to be important parameters controlling the phase diagram.

\section{Conclusion}

In summary, the glassy behavior we observe in the dynamics reveal a highly inhomogeneous system in a region of the phase diagram that is nominally a homogeneous disordered paramagnetic phase.  The NMR response probes the Fe spins through the hyperfine coupling, but it is the nematicity that drives the response of the system.  The disorder introduced by the dopants generates random strain fields, which couple to the nematicity and may contribute to the suppression of the nematic ordering temperature.   The nematic order parameter develops a complex fluctuating spatial landscape, with various domain sizes.    Future measurements under uniaxial strain  may significantly suppress the width of the distribution, and will provide an important avenue to investigate the dynamics in the glassy phase.  NMR studies of the dynamics under pressures up to 10-15 GPa in stoichiometric samples will also help to elucidate the role of disorder in suppressing the nematic phase.

\begin{acknowledgements}

We thank A. Thaler for assistance with initial sample growth, A. Estry for assistance with field calibration studies, as well as A. Benali, I. Fisher, S. Kivelson, E. Carlson and K. Dahmen for enlightening discussions. Work at UC Davis was supported by the NSF under Grant No.\ DMR-1005393. H.-J. G. acknowledges support by the Deutsche Forschungsgemeinschaft (DFG) through SPP1458 (Grants No. GR3330/2). Part of this  work  performed at the Ames Laboratory (PCC, SLB, SR) was supported by the U.S. Department of Energy, Office of Basic Energy Science, Division of Materials Sciences and Engineering. Ames Laboratory is operated for the U.S. Department of Energy by Iowa State University under Contract No. DE-AC02-07CH11358. A portion of this work was performed at the National High Magnetic Field Laboratory, which is supported by National Science Foundation Cooperative Agreement No.\ DMR-1157490, the State of Florida, and the U.S. Department of Energy.

\end{acknowledgements}

\bibliography{Ba122_nematic_glass_PRB}

\bibliographystyle{apsrev4-1}

\end{document}